\documentclass[11pt]{article}

\usepackage[a4paper,hmargin=1.72cm,vmargin=2.4cm]{geometry}
\usepackage{amsmath,amssymb,amsthm}
\usepackage{mathtools}
\usepackage{bbm}
\usepackage{hyperref}
\usepackage{caption}
\hypersetup{colorlinks=true,linkcolor=blue,citecolor=blue,urlcolor=blue}

\newcommand{\R}{\mathbb{R}}

\newcommand{\dd}{\,\mathrm{d}}

\begin{document}
\title{Win-score promotion gates in aggregator-routed RFQ markets:\\A two-tier stochastic control model}
\author{Alexander \textsc{Barzykin}\footnote{HSBC, 8 Canada Square, Canary Wharf, London E14 5HQ, United Kingdom, \texttt{alexander.barzykin@hsbc.com}.}}
\date{\today}

\maketitle

\vspace{-1.2em}
\abstract{
\noindent
We study market making in aggregator-routed RFQ markets where platform routing depends on slowly varying dealer performance scores.
We propose a two-tier stochastic control model that separates RFQ-level price competition from a macro routing layer:
tier A represents aggregator flow whose opportunity intensity is multiplied by a promotion gate driven by the dealer's win score,
while tier B captures background flow that is not gated and does not update the score.
RFQs arrive in multiple sizes and the dealer chooses a size-ladder of bid/ask offsets; conditional on winning, trades earn spread minus an
adverse selection correction and contribute to inventory risk.
The resulting Hamilton-Jacobi-Bellman equation admits a reduced Bergault-Gu\'eant operator form with explicit win/lose branches for the score on tier A.
Using the envelope-theorem argument, we express optimal controls through derivatives of the one-dimensional reduced Hamiltonians, yielding an interpretable mapping from optimal win probabilities to optimal offsets.
In the long-memory regime, we derive an adiabatic approximation that separates fast inventory dynamics from slow score dynamics.
A quadratic inventory ansatz and quadratic Hamiltonian expansion lead to a quasi-stationarity inventory-curvature scaling and a one-dimensional score drift field.
For steep (logistic) promotion gates, the score dynamics can exhibit fold bifurcations, bistability, and hysteresis, producing an endogenous ``campaign vs. harvest'' pattern in optimal quoting.
Numerical experiments confirm this behaviour and highlight the stabilizing role of background flow in maintaining inventory-mixing capacity even when the dealer is weakly promoted.
}

\vspace{7mm}


\vspace{5mm}

\section{Introduction}

Multi-dealer request-for-quote (RFQ) protocols are central to dealer-to-client execution in several over-the-counter (OTC) markets, including foreign exchange (FX). 
In an RFQ, a client requests prices from multiple liquidity providers (LPs) and either trades with one LP or does not trade at all. 
In modern FX workflows, RFQs are frequently routed through aggregators that consolidate quotes and automate selection, creating an interaction between {\em micro}-level price competition inside each RFQ and {\em macro}-level flow allocation rules that determine which LPs are invited to compete and how frequently.

On the modelling side, stochastic optimal control approaches to market making combine inventory management with quote-dependent order arrivals \cite{HoStoll1981,AvellanedaStoikov2008,GueantLehalleFernandezTapia2013,CarteaJaimungalRicci2014}. 
In OTC RFQ markets, an additional structural feature is size heterogeneity: an LP typically answers a menu of quotes for different notionals. 
This leads naturally to multi-size ladder controls and to Hamilton-Jacobi-Bellman (HJB) equations with non-local (jump) operators indexed by trade size \cite{BergaultGueant2019}.
Empirical and structural studies on multi-dealer-to-client (MD2C) platforms (notably in corporate bonds) explicitly model how hit probabilities depend on the competitive set and on the number of solicited dealers, highlighting that ``how many dealers are asked'' is a first-order primitive in RFQ economics \cite{FermanianGueantPu2016}.
More recently, causal interventions have been used  to study counterfactual pricing and revenue questions on MD2C platforms, bridging structural RFQ models with modern discriminative learning \cite{MarinArdanzaSabio2025}.

In FX, protocol design layer is particularly important. 
First, the presence of last look (a short post-trade window in which an LP may accept or reject) affects both spreads and selection \cite{OomenLastLook2017,CarteaJaimungalWalton2018}. 
Second, even without last look, the aggregator itself shapes adverse selection and incentives. 
Oomen \cite{OomenAggregator2017} argues that the ``winner's-curse'' in an aggregator can strengthen  as more LPs compete: the best displayed quote is more likely to be selected precisely in states that are unfavorable to the winning LP.
This provides an economic rationale for limiting the number of LPs exposed to each RFQ and for using routing and ranking rules rather than always inviting the full pool. 
Recent work on broker performance evaluation and selection highlights the practical importance of quantitatively ranking execution providers using intraday cost models \cite{EislerMuhle-Karbe2024}.
Related FX microstructure considerations, such as internalisation versus externalisation of client flow \cite{ButzOomen2019,BarzykinBergaultGueant2023} and informational risks that accompany competitive quoting \cite{BarzykinBergaultGueantLemmel2025}, further motivate modelling the interaction between client access, dealer incentives, and platform design.

A growing modern literature examines competition among liquidity providers and the extent to which it can be represented in tractable control or equilibrium frameworks. 
One approach treats competition in reduced form: a reference market maker optimises quotes while fill probabilities depend on differences versus exogenous competing quotes, yielding approximate closed-form solutions under linear-quadratic objectives \cite{BoyceHerdegenSanchez2024}. 
Other works move to explicit strategic settings (e.g.,\ Stackelberg or Nash structures), providing equilibrium characterisations for competitive liquidity provision \cite{ChilenjeEtAl2025,CarteaJaimungalSanchezBetancourt2024,DonnellyLi2025}.
A parallel strand uses reinforcement learning and agent-based models to study the emergence of tacit coordination in electronic markets and in MD2C-type dealer platforms, and the role of heterogeneity in mitigating such effects \cite{CarteaChangPenalva2022,CarteaChangMroczkaOomen2022,AssayagEtAl2024,ContXiong2024}.

Across these strands, two empirical facts stand out in aggregator-routed RFQ markets: (i) the competitive set is typically a {\em subset} of all eligible LPs and is often shaped by ranking, relationship, or platform-level ``top list + exploration'' logic; 
(ii) inclusion and rank depend on slowly varying performance scores (e.g., long-run win ratio, response quality), which in turn depend on how aggressively the LP quotes. 
While the RFQ micro-link is well captured by size-ladder market making \cite{BergaultGueant2019} and the macro design layer is motivated by aggregator economics \cite{OomenAggregator2017}, there is limited academic control literature that treats platform shortlisting driven by a slow win-score as an explicit state variable feeding back into the LP's future opportunity intensity.

This paper proposes a minimal two-scale model that links RFQ-level market making to aggregator routing via a slow win-score.
A key modelling choice is to start from the onset with two tiers of RFQ flow:
\begin{itemize}
\item an {\em aggregator} tier (A), whose opportunity intensity is multiplied by a gate $G(R)$ driven by the LP's score and whose outcomes update the score;
\item a {\em background} tier (B), representing residual client flow that is not gated by the platform score and does not update it.
\end{itemize}
This decomposition is motivated by practice: (a) only a subset of the firm's franchise is routed through ranking-based shortlists; (b) inventory risk is managed against all trading opportunities, not only the ranked stream.
Analytically, the background tier plays a stabilising role: it provides a baseline level of inventory mixing even when $G(R)$ is small, and it allows the model to generate a ``campaign vs. harvest'' trade-off in which the LP can temporarily sacrifice edge on tier A to improve future access while still managing inventory through tier B.

We derive a reduced HJB in a Bergault-Gu\'eant operator form, with explicit ``win'' and ``lose'' branches in the score for tier A, and standard jump operators for tier B.
Using an envelope-theorem argument, we express optimal controls through derivatives of scalar reduced Hamiltonians, giving a clean inverse mapping from optimal win probabilities to optimal offsets.
In the long-memory-EMA limit, the model separates into a fast inventory-control scale and a slow reputation-dynamics scale. 
A quadratic inventory ansatz and quadratic Hamiltonian expansion \`{a} la Bergault {\itshape{et al.}} \cite{BergaultEtAl2021} yield an LQ-like closure, stationary Riccati scaling for the inventory curvature, and a one-dimensional reduced dynamics for the score that admits phase transitions and hysteresis under steep (logistic) promotion gates.

\section{Model}

The mid-price $S_t$ is a martingale diffusion with volatility $\sigma > 0$, assumed constant over a finite operational horizon $[0, T]$ of the market maker.
The LP holds inventory $q_t$ and cash $X_t$.
RFQs arrive in multiple sizes $z\in \mathcal{Z} =\{z_1,\ldots,z_K\}$.
The LP chooses size- and side-dependent offsets to mid $\delta^{i,j}_z(t,q,R)$, where $i \in \{A, B\}$ denotes the tier (aggregator/background) and $j \in \{b,a\}$ denotes the side (bid/ask).\footnote{Positive offset means wider quote.}
Here $R$ is the long-term LP rating or score, defined below.
Given an RFQ opportunity of size $z$ and a quoted offset $\delta$, the LP wins with probability\footnote{We keep $p_z$ generic; logistic and structural erfc-type forms are natural candidates for numerical work.}
\begin{equation}
p_z(\delta)\in(0,1),\qquad p_z'(\delta)<0.
\end{equation}
Conditional on winning an RFQ of size $z$ at offset $\delta$, the expected mark-to-mid edge per unit size is modelled as $\delta+\mu_z(\delta)$, where $\mu_z(\delta)\le 0$ is an adverse-selection correction. 
To keep the analysis focused on win-score feedback, we model only latency-type adverse selection as an instantaneous expected edge correction conditional on winning the RFQ and do not include additional post-trade price jumps.
A simple benchmark is constant negative slippage $\mu_z(\delta)\equiv -\theta_z$, with $\theta_z>0$.

For each size $z$, RFQ opportunities arrive (per side) at baseline intensities $\Lambda_z^i>0$, assumed symmetric for simplicity.
Tier A opportunities are multiplied by a macro selection factor $G(R)\ge 0$:
\begin{equation}
\lambda_z^A(R)=\Lambda_z^A\,G(R), \qquad \lambda_z^B(R)=\Lambda_z^B .
\end{equation}
The gate $G(R)$ captures ranking-based shortlisting by an aggregator.
A stylized ``top list + exploration'' mechanism is: the  $N_{\mathrm{top}}$ highest-score LPs are always invited (promotion), and the remaining  $N_{\mathrm{select}} - N_{\mathrm{top}}$ invites are sampled at random from the rest (exploration).
Thus, an LP's long-run score influences the probability of being invited and, therefore, the RFQ opportunity intensity.
The full environment is a game: win probabilities depend on the cross-sectional distribution of competitor quotes.
In the mean-field regime, this can be represented as a single-LP stochastic control with a monotone $G(R)$.

Let $w\in\{0,1\}$ denote the win indicator on a tier A RFQ opportunity. 
We update the score (EMA of wins) as follows:
\begin{equation}
R\mapsto R_\pm(R):=(1-\alpha)R+\alpha w,
\end{equation}
where $\alpha \in (0, 1)$ and $\pm$ correspond to $w=1$ and 0, respectively.\footnote{One can introduce size-weighted updates with $w_z = z /z_{\max}$.}
Tier B outcomes do not update $R$.
When $\alpha$ is small, the score evolves slowly relative to inventory dynamics.

We assume a standard linear wealth objective with quadratic inventory penalties \cite{CarteaJaimungalRicci2014}:
\begin{equation}
\sup_{\delta}\;
\mathbb{E}\Big[
X_T + q_T S_T - \frac{\eta}{2}q_T^2 - \frac{\gamma}{2}\sigma^2 \int_0^T q_t^2\,\dd t
\Big],
\end{equation}
with $\eta,\gamma\ge 0$.

\subsection{Reduced HJB}

Let $V(t,x,q,S,R)$ denote the value function. 
The standard reduction $V(t,x,q,S,R)=x + qS + v(t,q,R)$ eliminates explicit dependence on $S$, with terminal condition
$v(T,q,R)=-\frac{\eta}{2}q^2$.
Define the lose-branch decrement and win-and-trade increments for tier A:
\begin{align}
\mathcal{L}(t,q,R) &:= v(t,q,R_-(R)) - v(t,q,R),\\
\Delta^{A,j}_z(t,q,R) &:= v(t,q\pm z,R_+(R)) - v(t,q,R) ,
\end{align}
where plus corresponds to bids and minus to asks, respectively.
Then for each size $z$, define the tier-A Bergault-Gu\'eant operator
\begin{equation}
\mathcal{H}^A_z(\Delta) = 
\sup_{\delta\in\mathbb{R}} \Big\{ p_z(\delta)\big(z(\delta+\mu_z(\delta))+\Delta\big) + \big(1-p_z(\delta)\big)\mathcal{L} \Big\}
= \mathcal{L} + \sup_{\delta\in\mathbb{R}} p_z(\delta)\Big(z(\delta+\mu_z(\delta))+\Delta-\mathcal{L}\Big).
\end{equation}
The first equality makes explicit the win/lose mixture: with probability $p_z(\delta)$ the dealer wins and trades, receiving spread and moving to $(q\pm z, R_+)$, while with probability $1 - p_z(\delta)$ the dealer loses and only incurs the score downgrade $(q, R_-)$.
Since tier B does not update $R$, define
\begin{equation}
\Delta^{B,j}_z(t,q,R) := v(t,q\pm z,R) - v(t,q,R) ,
\end{equation}
\begin{equation}
\mathcal{H}^B_z(\Delta) = \sup_{\delta\in\mathbb{R}} p_z(\delta)\Big(z(\delta+\mu_z(\delta))+\Delta\Big).
\end{equation}

Assuming symmetric primitives across bid/ask (same $\Lambda_z,p_z,\mu_z$), the reduced HJB is given by
\begin{equation}
0=\partial_t v - \frac{\gamma}{2}\sigma^2 q^2
+\sum_{z\in\mathcal{Z}, i\in\{A,B\}, j\in\{b,a\}} \lambda_z^i(R) \mathcal{H}_z^i\big(\Delta^{i, j}_z\big).
\end{equation}

\subsection{Envelope theorem and optimal controls}

Define the reduced scalar Hamiltonian
\begin{equation}
H_z(x):=\sup_{\delta\in\R}\;p_z(\delta)\Big(z(\delta+\mu_z(\delta))+x\Big).
\label{eq:reduced_hamiltonian}
\end{equation}
Then $\mathcal{H}_z^A(\Delta)=\mathcal{L}+H_z(\Delta-\mathcal{L})$ and $\mathcal{H}_z^B(\Delta)=H_z(\Delta)$.
Assume the maximiser $\hat \delta_z(x)$ in \eqref{eq:reduced_hamiltonian} is unique and interior. 
The envelope theorem yields
\begin{equation}
H'_z(x)=p_z\big(\hat \delta_z(x)\big)\in(0,1),
\qquad
\hat \delta_z(x)=p_z^{-1}\big(H'_z(x)\big).
\end{equation}
Hence optimal win probabilities and offsets can be written directly in terms of $v$.
\begin{equation}
\hat \delta_z^{i,j}(t, q, R) = p_z^{-1} \big(\hat y_z^{i,j}(t, q, R)\big), \qquad
\hat y_z^{i,j}(t, q, R) = H'_z\big(x_z^{i,j}(t,q,R)\big),
\end{equation}
where
\begin{equation}
x_z^{A,j}(t, q, R) := \Delta_z^{A,j}(t, q, R) - \mathcal{L} (t, q, R), \qquad
x_z^{B,j}(t, q, R) := \Delta_z^{B,j}(t, q, R).
\end{equation}
This representation is convenient analytically and numerically: the controls are recovered from derivatives of scalar Hamiltonians, and the dependence on the non-local value function enters only through the one-dimensional arguments $x_z^{i,j}$.

\section{Adiabatic quadratic approximation}

When $\alpha$ is small, score changes are slow relative to inventory dynamics.
Following the two-scale philosophy used in our earlier last look reputation study \cite{Barzykin2026LastLook}, we separate:
(i) a fast inventory problem conditional on quasi-static $R$, and (ii) a slow score dynamics driven by the induced average win probability on tier A.

\subsection{Fast scale: small-$\alpha$ expansion and quadratic ansatz}

For smooth $v$ we have, to the first order of $\alpha$,
\begin{equation}
v(t, q, R_+) \approx v(t, q, R) + \alpha(1-R)v_R(t, q, R), \qquad v(t, q, R_-) \approx v(t, q, R) - \alpha v_R(t, q, R).
\end{equation}
Therefore, tier-A continuation admits the decomposition
\begin{equation}
x_z^{A,j}(t, q, R) = v(t, q\pm z, R) - v(t, q, R) + \alpha \big((1-R) v_R (t, q\pm z, R) + R v_R(t, q, R)\big) + O(\alpha^2),
\end{equation}
where we can clearly distinguish the fast inventory term and the slow score term.
The $O(\alpha)$ term acts as an additive reward for winning, a slowly varying ``prize'' for reaching higher $R$.
It is proportional to the marginal value $v_R$ and is strongest where $v_R$ is large,
which in turn is tied to the steepness of the routing gate $G(R)$.
Tight quoting can be optimal even when locally unprofitable because it increases future tier-A access through $R$.
Tier B gaps are purely fast: $x_z^{B,j}(t, q, R) = v(t, q\pm z, R) - v(t, q, R)$.

In the spirit of \cite{BergaultEtAl2021}, we assume a quadratic ansatz in inventory\footnote{No linear term due to bid/ask symmetry.}
\begin{equation}
v(t, q, R) \approx -\frac12 A(t,R)\,q^2 + B(t,R),
\end{equation}
with $A(T, R) = \eta$, $B(T, R) = 0$ (since terminal condition has no $R$-dependent constant term.
Then, for small $q$, $v(t, q\pm z, R) -v(t, q, R) \approx -\frac{1}{2}A(t,R) z (z \pm 2q)$ and $v_R(t, q, R) \approx B_R(t, R)$,
so the continuation gaps become
\begin{equation}
x_z^{A,j} (t, q, R) \approx -\frac{1}{2}A(t, R) z (z \pm 2q) + \alpha B_R(t, R), \qquad
x_z^{B,j} (t, q, R) \approx -\frac{1}{2}A(t, R) z (z \pm 2q).
\end{equation}

We further expand each $H_z$ around $x^i_{0, z}:=x_z^{i, j}(t, 0 ,R)$:
\begin{equation}
H_z(x)\approx H_z(x_{0, z}^i) + y_{0, z}^i(t, R)\,(x - x_{0, z}^i) + \frac12 k_z^i(t, R)\,(x - x_{0, z}^i)^2,
\end{equation}
where $y_{0,z}^i := H'_z(x_{0, z}^i)$ and $k_z^i := H''_z(x_{0, z}^i)\ge 0$.
Thus, to leading order around $q=0$,
\begin{equation}
\hat y^{i,j}_z(t, q, R)\approx y_{0,z}^i(t, R)\mp k_z^i(t, R)\,z\,q\,A(t, R), \quad i \in \{A,B\}.
\label{eq:adiabatic_quadratic_win_p}
\end{equation}
Offsets follow by $\delta = p^{-1}(y)$.
This shows explicitly how inventory tilts bid/ask win probabilities via the familiar $z q A(t, R)$ factor, while tier A carries an additional $O(\alpha)$ shift through $x_{0,z}^A$.

\subsection{Ergodic curvature scaling}

On the fast scale, $A(t,R)$ relaxes quickly to a quasi-stationary value at fixed $R$.
Matching $q^2$ coefficients in the reduced HJB under the quadratic Hamiltonian expansion\footnote{Details follow the standard Bergault-Gu\'eant calculations.} yields a Riccati-type relation in which the stabilizing contribution comes from both tiers.
In a stationary inner approximation, one obtains the scaling 
\begin{equation}
A_\infty(R)\approx \sqrt{\frac{\gamma\sigma^2}{2\xi(R)}},
\qquad
\xi(R):=\sum_{z\in\mathcal{Z}, i \in \{A,B\}}\lambda_z^i\,k_z^i(R)\,z^2.
\end{equation}
This expression emphasizes the intuitive point: higher flow makes inventory easier to manage (smaller curvature).\footnote{One can reproduce familiar optimal controls in the absence of score feedback by using riskless baseline $H_z'(0) = p_z(\delta_z^0)$.}

When $k_z^A \approx k_z^B \approx k_z$ (a convenient closure near $q = 0$), the only strong $R$-dependence in $A_\infty$ is through the intensity scale $\sum_{z,i} \lambda_z^i(R)k_z z^2$.
In particular, the presence of tier B prevents the singular $A_\infty \propto 1/\sqrt{G(R)}$ behavior that arises when {\em all} flow is gated: even at $G(R) \approx G_{\min}$ curvature remains bounded thanks to baseline mixing from the background tier.

\subsection{Slow score dynamics and phase transitions}

A convenient summary of the slow dynamics is through the expected drift of $R$.
Let $N_t^A$ denote the tier-A RFQ counting process (both sides) with intensity
\begin{equation}
\lambda_A^{\rm RFQ}(R) = \sum_{z\in\mathcal{Z}, j \in \{b,a\}}\lambda_z^A(R) = 2G(R)\sum_{z\in\mathcal{Z}}\Lambda_z^A,
\end{equation}
and let $w_k \in \{0, 1\}$ be the win indicator on the $k$-th tier-A RFQ.
The EMA recursion in event time is $R_{k+1} = (1-\alpha)R_k + \alpha w_k \Longrightarrow \Delta R_k = \alpha (w_k - R_k)$.
Passing to continuous time and taking conditional expectation leads to the drift approximation
\begin{equation}
\label{eq:slow_score_ode}
\dot R \approx \alpha\,\lambda_A^{\rm RFQ}(R)\,\big(\bar y_A(R) - R\big),
\end{equation}
where $\bar y_A(R)$ is the tier-A intensity-weighted win probability near fast equilibrium ($q \approx 0$): 
\begin{equation}
\bar y_A(R) := \frac{\sum_{z\in\mathcal{Z}} \Lambda_z^A\, y_{0,z}^A(R)}{\sum_{z\in\mathcal{Z}} \Lambda_z^A}
= \frac{\sum_{z\in\mathcal{Z}} \Lambda_z^A\, p_z\big(\hat \delta_z^A(0, R)\big)}{\sum_{z\in\mathcal{Z}} \Lambda_z^A}.
\end{equation}

Equilibria satisfy $\bar y_A(R) = R$, and stability is determined by $\bar{y}'_A(R) - 1$.
A phase transition corresponds to the emergence of three fixed points (two stable separated by one unstable), producing bistability and hysteresis.
The background tier affects this picture indirectly through the fast-scale coefficients $A_\infty(R)$ and hence $y_{0,z}^A(R)$, while the score dynamics itself is driven only by tier A.

\subsection{Logistic promotion gate and minimal phase-portrait closure}

Assume a logistic gate, capturing shortlisting:
\begin{equation}
G(R) = G_\textrm{min} + \Delta G\, u, \qquad u := \left(1 + e^{-\beta (R-R_0)}\right)^{-1},
\end{equation}
with steepness $\beta$, midpoint $R_0$ and $\Delta G = G_{\max} - G_{\min}$.
Then $G'(R) = \beta \Delta G\, u (1-u)$ and
\begin{equation}
R(u) = R_0 + \frac{1}{\beta} \log \frac{u}{1-u}.
\end{equation}

To obtain a compact scalar phase portrait, one may postulate a reduced-form closure for $\bar y_A(R)$:
\begin{equation}
\label{eq:win_p_closure}
\bar y_A(R) \approx \bar y_\star + \mathcal{A}\Big(\frac{1}{\sqrt{\xi(R_0)}}-\frac{1}{\sqrt{\xi(R)}}\Big) + \alpha \mathcal{B}\,G'(R),
\end{equation}
where $\bar y_\star := \sum_z \Lambda_z^A H'_z(0) / \sum_z \Lambda_z^A$ is the baseline ``riskless'' win rate, the second term encodes the Riccati scaling $A_\infty(R) \propto 1/\sqrt{\xi(R)}$, and the last term captures the slow ``promotion wedge'' driven by $v_R$ and concentrated where the gate is steep.
Constants $\mathcal{A}, \mathcal{B} > 0$ encode microstructure features (the local curvature $p_z$, slippage, and ladder composition) and can be estimated from the quadratic expansion.

Define $F(R) := \bar y_A(R) - R$.
To make the fold structure explicit under a logistic gate, it is convenient to parametrize $R$ via $u \in (0, 1)$.
Assume for simplicity that the microstructure curvatures are frozen near the fast equilibrium, $k_z^A \approx k_z^B \approx k_z$, and define $\xi_i := \sum_z \Lambda_z^i k_z z^2$.
Then $\xi(R) = \xi(u) = \xi_B + (G_{\min} + \Delta G\, u) \xi_A$ and \eqref{eq:win_p_closure} yields a scalar fixed-point equation
\begin{equation}
F(u)=
\bar y_\star
+ \mathcal{A}\Big(\frac{1}{\sqrt{\xi_0}}-\frac{1}{\sqrt{\xi_B + (G_{\min}+\Delta G\,u)\xi_A}}\Big)
+ \alpha \mathcal{B}\,\Delta G\,\beta\,u(1-u)
- \Big(R_0+\frac{1}{\beta}\log\frac{u}{1-u}\Big)=0,
\end{equation}
where $\xi_0 = \xi(R_0)$.
Fold points (hysteresis boundaries) satisfy in addition
\begin{equation}
\frac{\dd F}{\dd u}=
\mathcal{A}\,\frac{\Delta G \xi_A}{2}\,\big(\xi_B + (G_{\min}+\Delta G u)\xi_A\big)^{-3/2}
+
\alpha \mathcal{B}\,\Delta G\,\beta\,(1-2u)
-
\frac{1}{\beta}\Big(\frac{1}{u}+\frac{1}{1-u}\Big).
\end{equation}

At least one fixed point is guaranteed since $\bar y_A(R) \in (0, 1)$ implies $F(0) > 0$ and $F(1) < 0$.
For sufficiently steep gates, the only region where $F$ can overcome the linear $-R$ term is near $R_0$, where $G'(R)$ is maximal.
For sufficiently strong score feedback (large $\alpha$ and/or steep $\beta$), the bell-shaped campaign term can generate two additional fixed points through a fold bifurcation, producing a hysteresis loop.
In this two-tier setting, the ``permanent'' low-score fixed point typically lies near the low-gate regime $G(R) \approx G_{\min}$ but remains quantitatively influenced by the baseline curvature $\xi(R)$ contributed by the background tier.

\section{Numerical examples.}

As an illustration, consider a standard size-ladder of $z = (1, 2, 5, 10)$ M notional with baseline RFQ intensities 
$\Lambda_z^B = (1000, 800, 600, 400)$~day$^{-1}$ for the background tier and $\bar \Lambda_z^A = (0, 0, 0, 50)$~day$^{-1}$ for the gated aggregator tier.
Individual clients often tend to have a narrow size distribution, and this is exactly the case we are capturing here.
Win probability is taken to be of a standard sigmoid shape,
\begin{equation}
p_z(\delta) = \left(1 + e^{\kappa_z (\delta - \bar \delta_z)} \right)^{-1} ,
\end{equation}
with $\kappa_z = (5.0, 4.5, 4.0, 3.5)$ bp$^{-1}$ and $\bar \delta_z = (0.3, 0.4, 0.5, 0.6)$ for both tiers.
Here bp stands for basis points.\footnote{This implies GBM while we deal here with simple Brownian motion. 
The difference is negligible in FX market making due to short trading horizons.}
A constant latency slippage of $\theta_z = 0.2$ bp is assumed across all sizes and tiers.
We also assume a daily volatility of $100$~bp and a risk aversion coefficient of $\gamma = 10^{-3}$~bp$^{-1}$~M$^{-1}$.
This set of parameters corresponds to a daily turnover of $ \approx 2$ billion notional and a top-of-book spread of $\approx 1$~bp.
The default gate parameters are $G_{\min} = 0.2$, $R_0 = 0.6$, $\beta = 40$ and the EMA memory parameter is $\alpha = 0.01$, unless specified otherwise.

\begin{figure}[h]
\centering
\includegraphics[width=0.618\columnwidth]{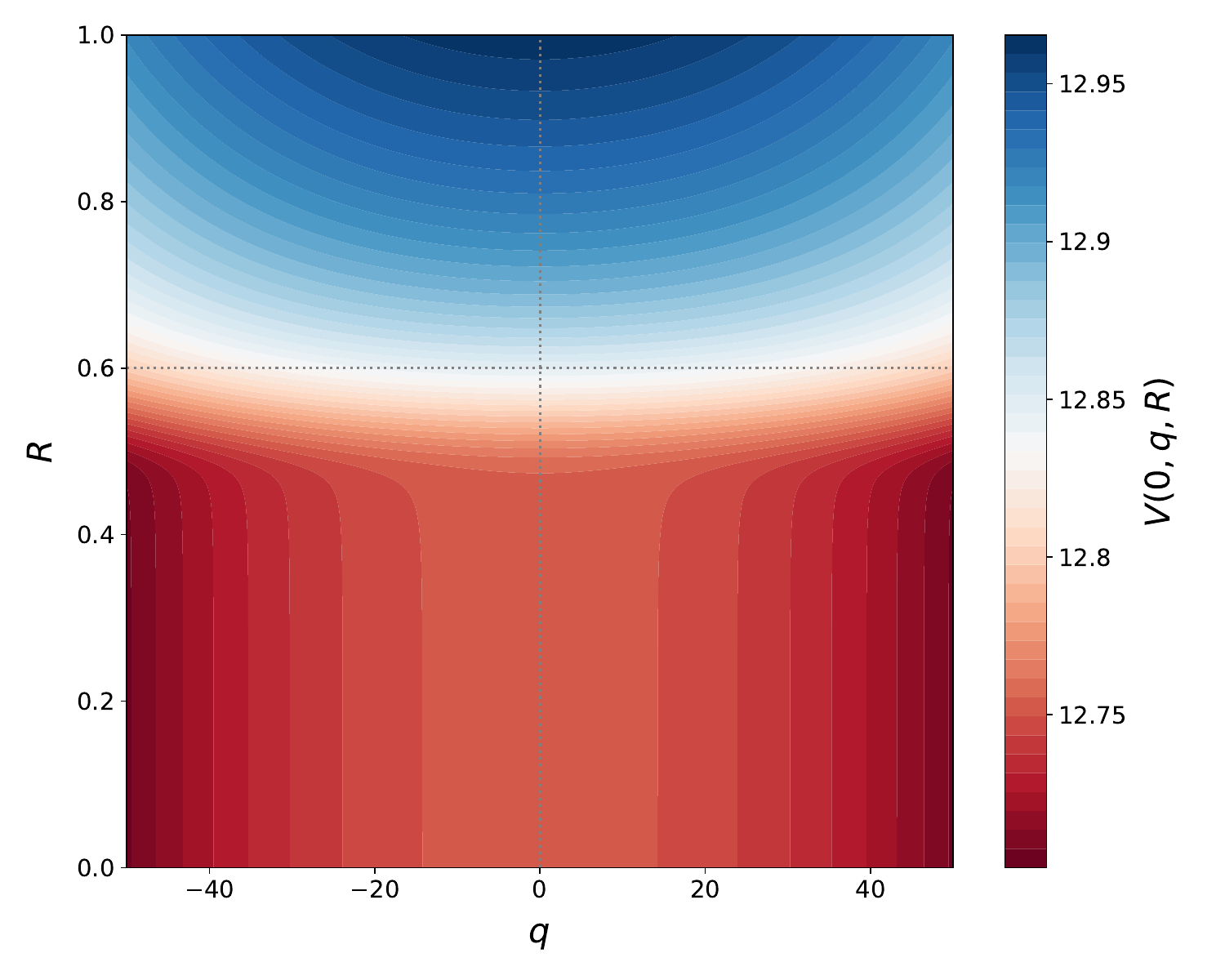}
\caption{
Heatmap of the stationary value function $v(0, q, R)$ for a default parameter set defined in the text.
Horizontal dotted line corresponds to $R_0$.
}
\label{value_function}
\end{figure}

Numerical solution of the HJB equation can be challenging, because on one hand, the time step $dt$ has to be sufficiently small for the Euler backward step to be valid, 
while on the other hand, the time horizon $T$ has to be sufficiently long to cover score relaxation, leading to a large number of steps and increased precision requirements.
For $\alpha = 0.01$, EMA half-life is $\log(0.5)/\log(1-\alpha) \approx 70$ trades, so at the minimum gate value of 0.2 this is equivalent to $\approx 3.5$ days for tier A, 
while the timestep has to be well below $1/\sum_z \Lambda_z \approx 30$ s, which brings the number of steps to $\gtrsim 10^6$.
We can take advantage of timescale separation and choose $\mathcal{T}$ long enough to cater for the inventory risk relaxation alone.
Then on time interval $[0, \mathcal{T}]$ we solve HJB with terminal condition $v(\mathcal{T}, q, R) = -\frac{1}{2}\eta q^2 + e^{-\rho\mathcal{T}} \Phi(R)$, where $\rho$ is the discount rate and $\Phi(R)$ is the continuation term that propagates the slow score incentives across intervals.
Stationarity across $\mathcal{T}$-intervals implies $\Phi(R) = v(0, 0, R)$.
So $\Phi$ is determined by a fixed point: the terminal condition depends on $\Phi$ and the start-of-interval value regenerates $\Phi$.
This approach is similar in spirit to stationary HJB, and also mimics business cycle. 
Technically, we start with $\Phi^{(0)}(R) = 0$, then for each $n$ solve interval HJB with terminal condition
\begin{equation}
v^{(n)}(\mathcal{T}, q, R) = -\frac{1}{2}\eta q^2 + e^{-\rho \mathcal{T}} \Phi^{(n)}(R)
\end{equation}
and update
\begin{equation}
\Phi^{(n+1)}(R) = (1-\zeta)\Phi^{(n)}(R) + \zeta v^{(n)}(0, 0, R)
\end{equation}
until the absolute increment is sufficiently small.
Here $\zeta$ introduces optional damping.
This way we constrain PDE horizon to a managable number of steps but at the same time capture multi-day score incentives.
Practically, to reduce the number of iterations we also apply Anderson acceleration \cite{WalkerNi2011}, which uses the last $m$ iterates to build a better next iterate by choosing a linear combination that  minimizes the residual with ridge regression.
Numerical calculations in this paper are based on the block interval of $\mathcal{T} = 0.05$ days with $10^4$ time steps ($dt \approx 0.5$ s).

\begin{figure}[h]
\centering
\includegraphics[width=0.618\columnwidth]{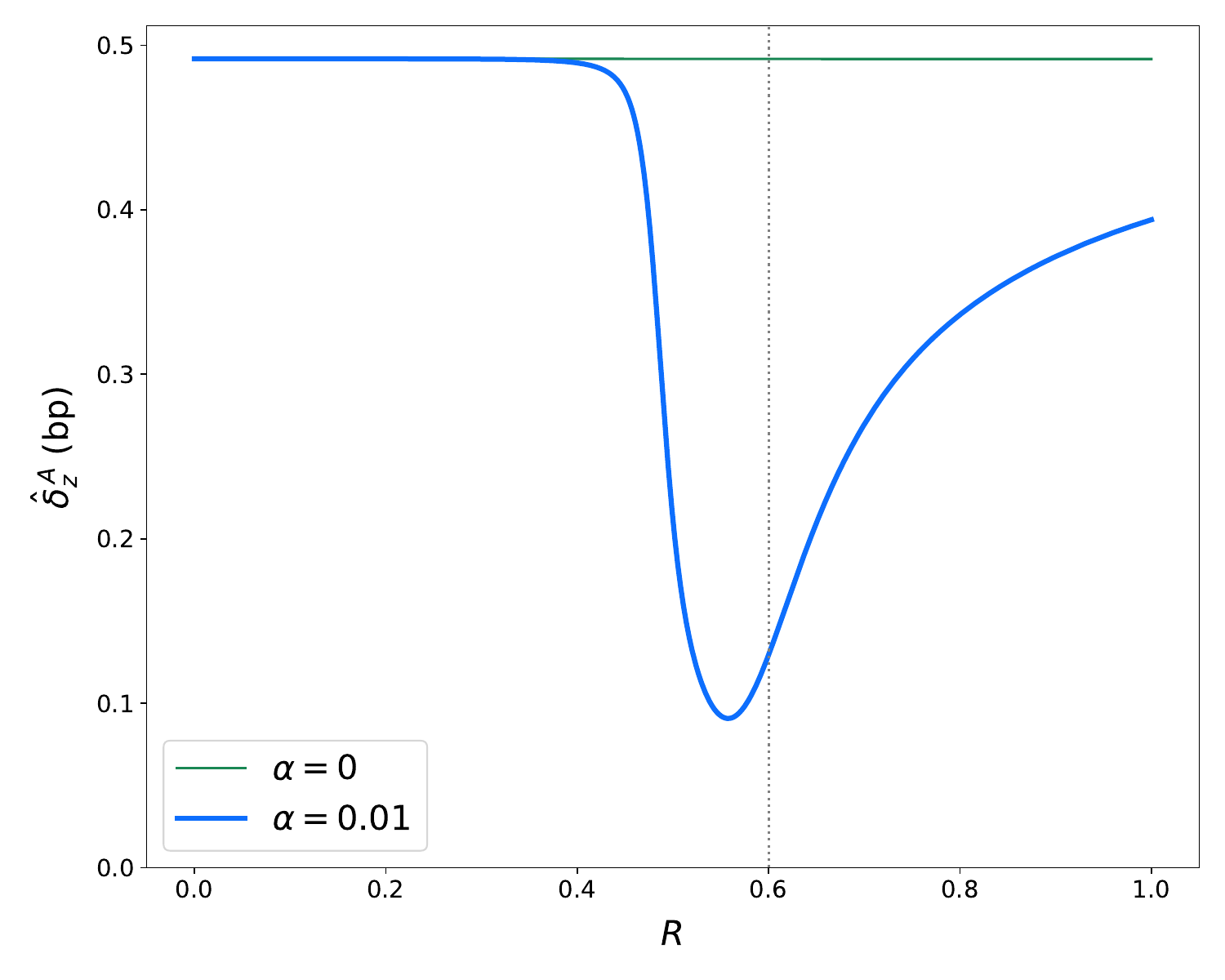}
\caption{
Tier A optimal quote offset $\hat \delta_z^A$ for $z = 10$ at zero inventory as a function of score $R$ for $\alpha = 0.01$ (feedback) and $\alpha =0$ (no feedback).
Other parameters are defined in the text.
Due to bid/ask parameter symmetry, side index is omitted.
Vertical dotted line corresponds to $R_0$.
}
\label{offset_R}
\end{figure}

\begin{figure}[!h]
\centering
\includegraphics[width=0.618\columnwidth]{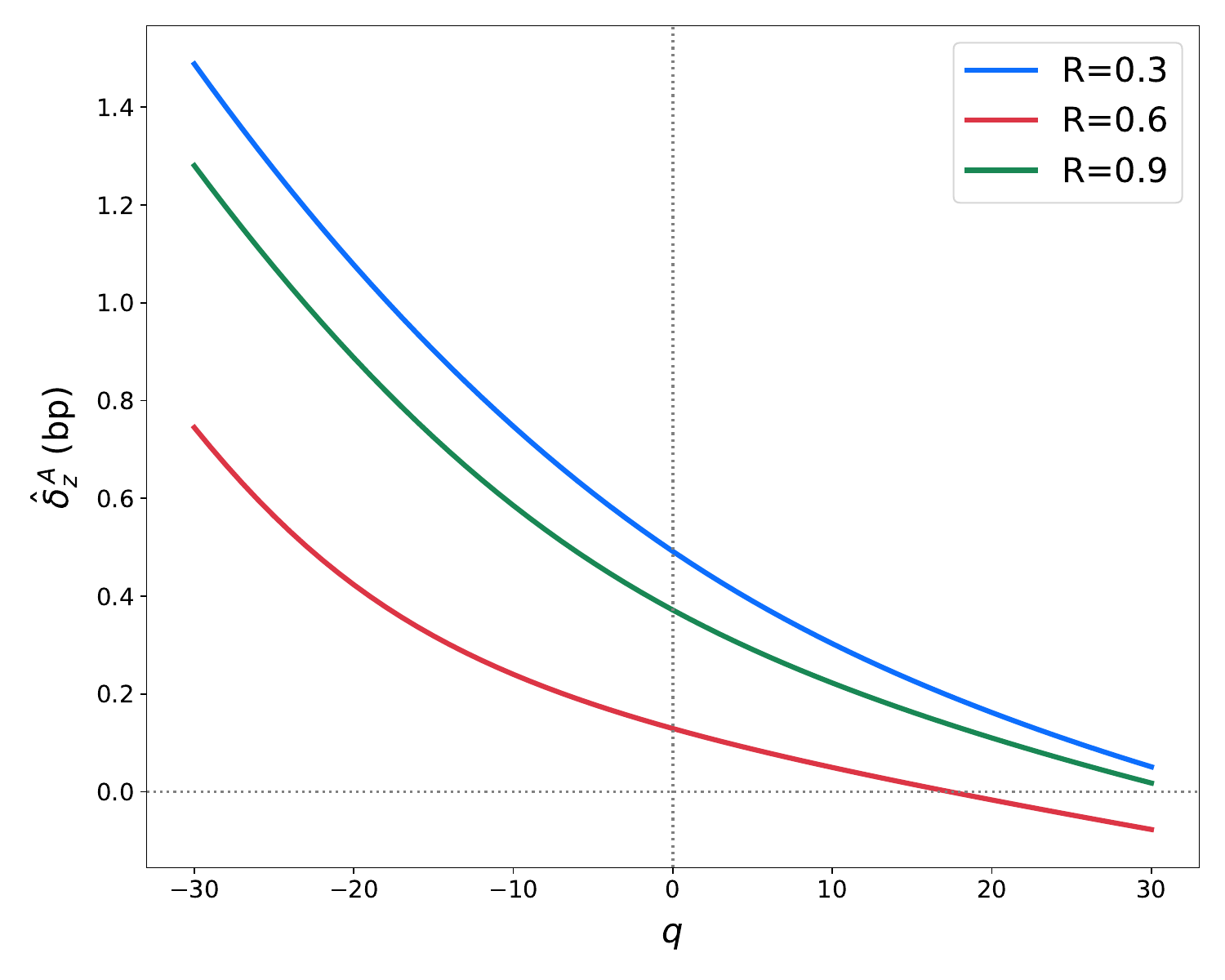}
\caption{
Tier A optimal quote offset $\hat \delta_z^A$ for $z = 10$ as a function of inventory $q$ for several values of score $R$.
Parameters are defined in the text.
Due to bid/ask parameter symmetry, side index is omitted.
}
\label{offset_q}
\end{figure}

Figure \ref{value_function} illustrates the heatmap of the stationary value function $v(0, q, R)$.
Gradients related to the inventory and score management are clearly visible.
Figure \ref{offset_R} demonstrates a dip in quote offset for tier A at zero inventory in the vicinity of $R_0$ -- the signature of campaigning.
Above $R_0$, the dealer would like to ``harvest'' and widens spread, but clearly not fully back to the level without feedback ($\alpha = 0$).
Widening too much immediately lowers the win rate $\bar{y}_A(R)$, and when $\bar{y}_A(R) < R$, score decay is accelerated.
If the gate is steep, the downside of decaying across the gate is large, so the optimal policy can remain defensive even at high $R$.
In other words, high score is not a riskless asset and maintenance costs.

Figure \ref{offset_q} shows the dependence of the quote offset for tier A as a function of inventory $q$ for several levels of score $R$.
Campaigning around $R_0$ is clear across inventories.
Note that for $R = 0.3$, gated offset matches not just the myopic value without feedback ($\alpha = 0$) but also the offset for the background tier B, same size $z = 10$.
This is because we have selected the same parameters for the win probability $p_z(\delta)$ for both tiers.

\begin{figure}[h]
\centering
\includegraphics[width=0.618\columnwidth]{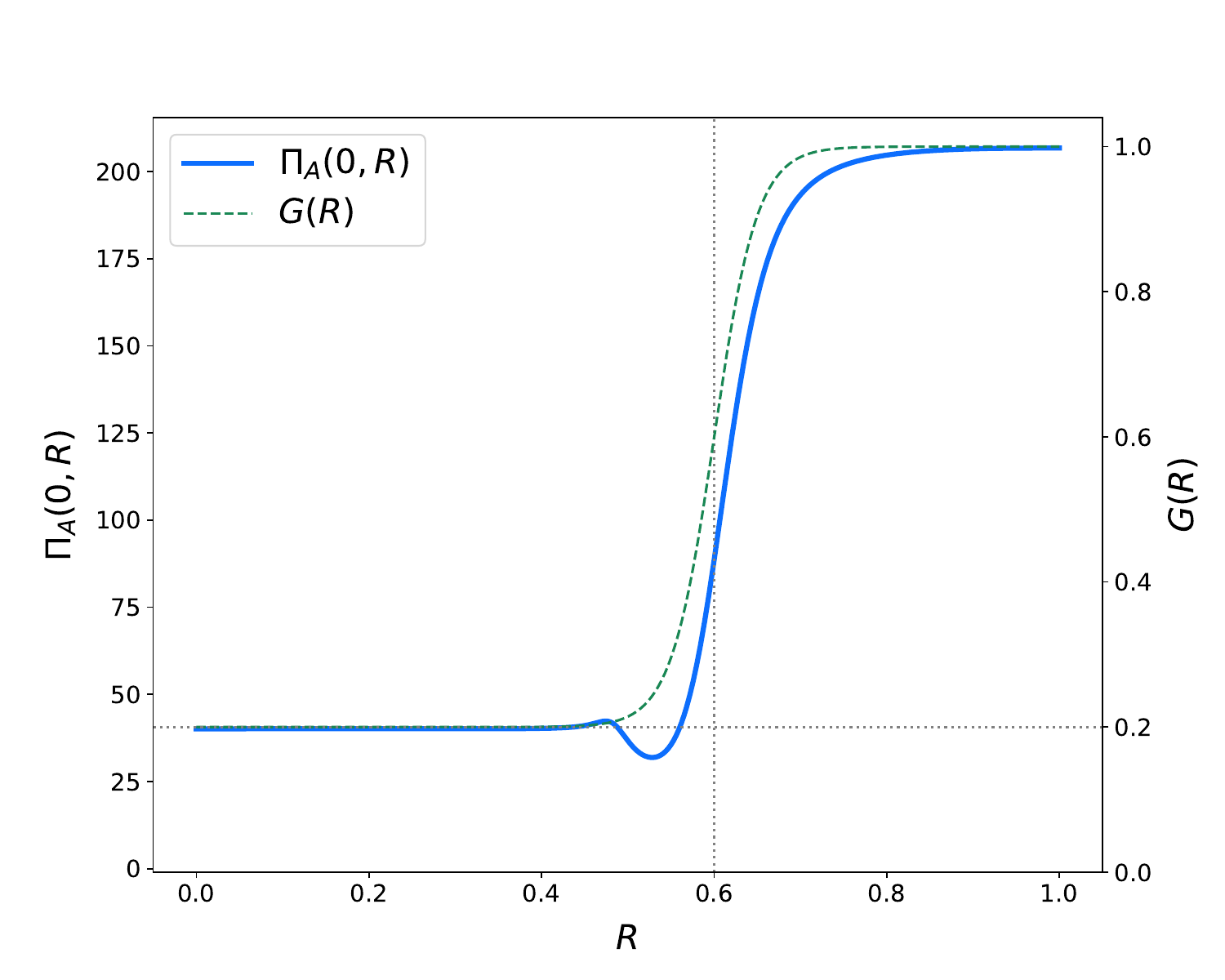}
\caption{
Logistic gate $G(R)$ (dashed line) and instant PnL of the gated tier at zero inventory $\Pi_A(0, R)$ (solid line) as functions of score $R$.
Parameters are defined in the text.
Vertical dotted line corresponds to $R_0$.
}
\label{pnl_R}
\end{figure}

\begin{figure}[!h]
\centering
\includegraphics[width=0.618\columnwidth]{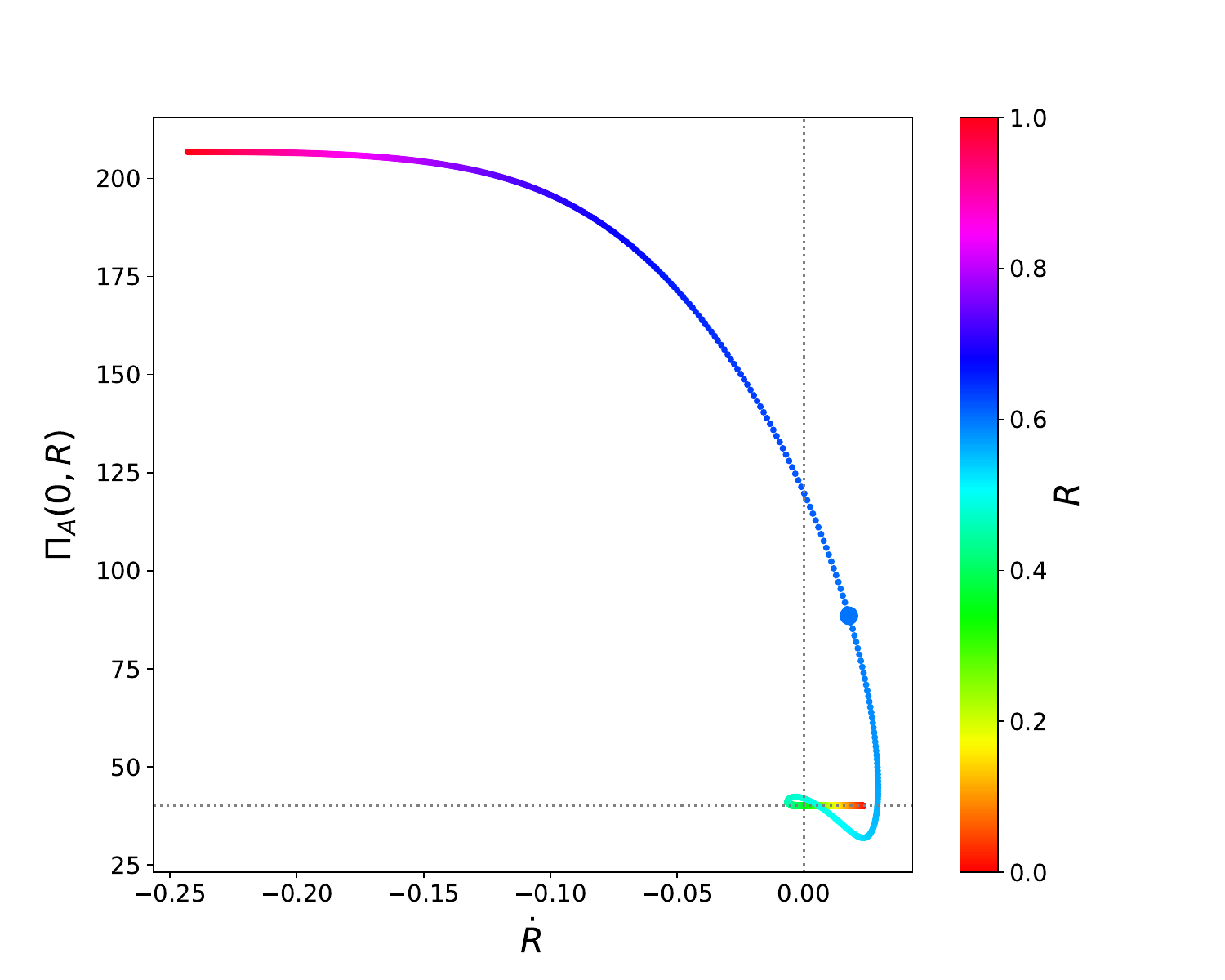}
\caption{
Instant PnL of the gated tier at zero inventory $\Pi_A(0, R)$ as a function of score drift $\dot R$ color coded by the value of $R$.
Parameters are defined in the text.
Larger circle on the curve corresponds to $R_0$.
Horizontal dotted line corresponds to $\Pi_A(0, 0)$.
}
\label{pnl_Rdot}
\end{figure}

The signature of campaigning and harvesting becomes even more dramatic in instantaneous PnL, which can be defined as
\begin{equation}
\Pi_A(q, R) = 2 G(R) \sum_z \Lambda_z^A p_z(\hat \delta_z^A) z ( \hat \delta_z^A - \theta_z)
\end{equation}
for the gated tier, where we have dropped side dependence due to symmetry and summed up to get the factor of 2.
Here $\Pi_A(q, R)$ should be interpreted as the instantaneous edge-capture rate (per unit time) from tier A only, conditional on the state $(q,R)$ and evaluated at the optimal tier-A offsets; it excludes the running inventory penalty term and any contribution from tier B.
Figure \ref{pnl_R} illustrates the dependence of the instantaneous PnL on the score along with the shape of the gate for comparison.
During campaigning right below $R_0$, PnL drops, but then rises sharply to enjoy harvesting.
The dependence of PnL on score drift $\dot R$ is also very representative, as shown in Figure \ref{pnl_Rdot}.
Campaign is characterized by positive $\dot R$ and low PnL while harvesting corresponds to negative $\dot R$ and high PnL.

\begin{figure}[h]
\centering
\includegraphics[width=0.618\columnwidth]{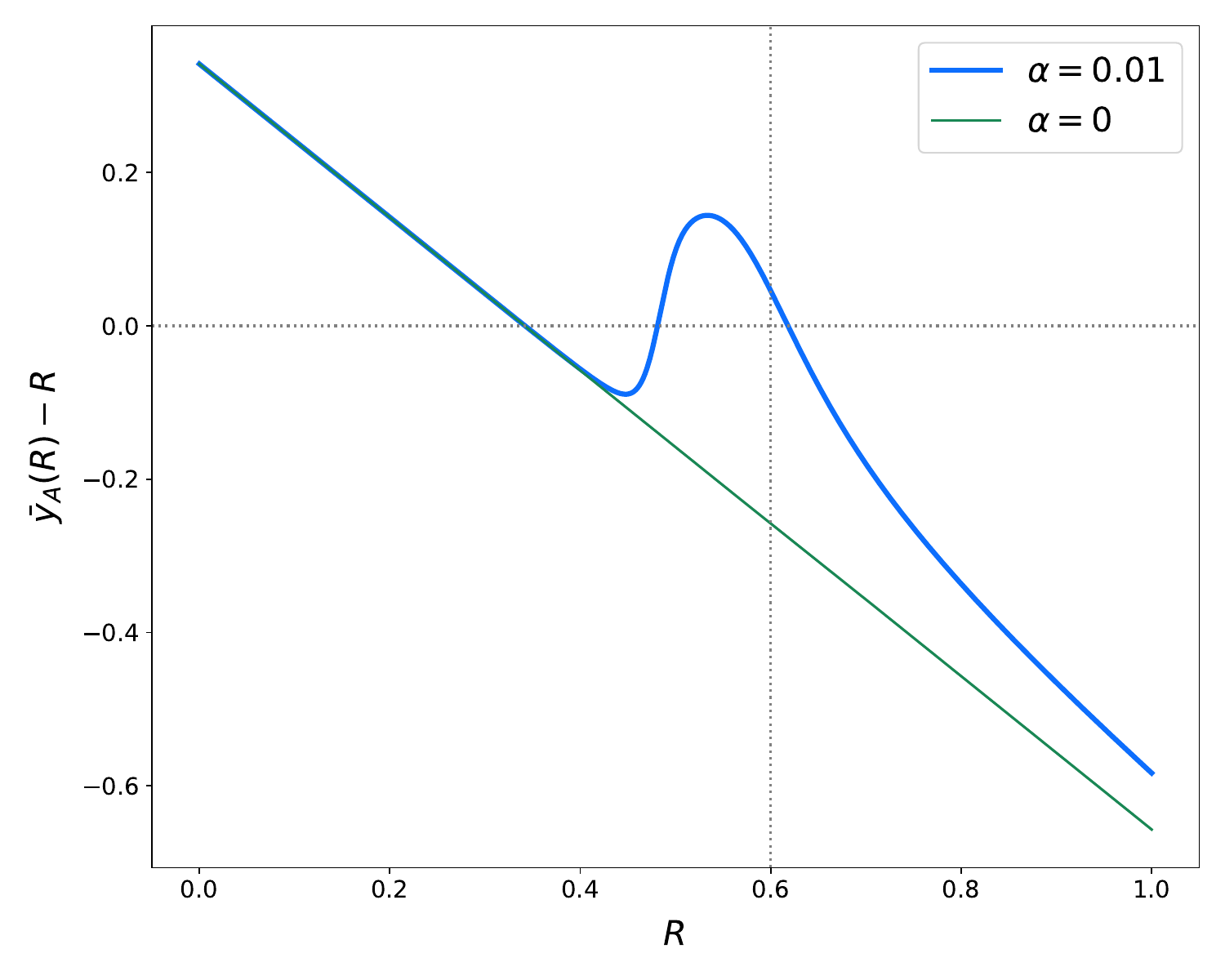}
\caption{
Score phase portrait for $\alpha = 0.01$ (feedback) and $\alpha =0$ (no feedback).
Other parameters are defined in the text.
Vertical dotted line corresponds to $R_0$.
}
\label{drift}
\end{figure}

\begin{figure}[h]
\centering
\includegraphics[width=0.618\columnwidth]{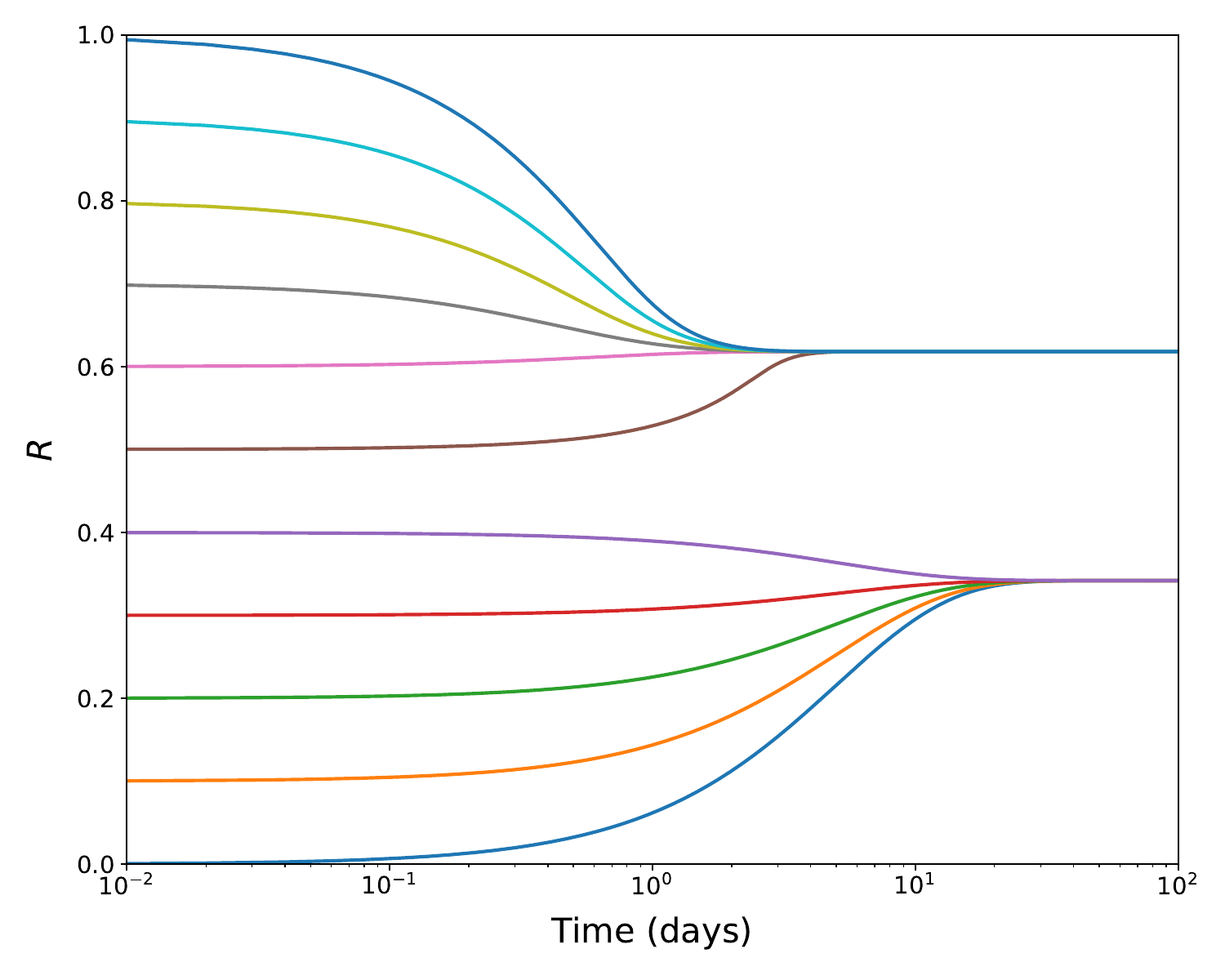}
\caption{
Slow score relaxation for different initial conditions demonstrating two stable basins.
Parameters are defined in the text.
}
\label{relaxation}
\end{figure}

Phase portrait of the score can exhibit a very interesting behavior typical of a critical system, as demonstrated in Figure \ref{drift}.
While the myopic model with $\alpha = 0$ exhibits a single fixed point, the score feedback can give rise to bistable behavior with three fixed points and two basins of attraction.
Figure \ref{relaxation} illustrates the mean-field relaxation trajectories (obtained by numerically solving the slow score ODE \eqref{eq:slow_score_ode} with optimal controls) for different initial conditions.
Below the lower fold, campaigning is not worth it (the gate is too far, the prize is too small).
Inside the bistable region, a sufficiently aggressive campaign can push $R$ across the unstable fixed point into the promoted basin.
Above the upper fold, harvesting becomes optimal because the dealer remains promoted even with wider quotes.

\begin{figure}[!h]
\centering
\includegraphics[width=0.618\columnwidth]{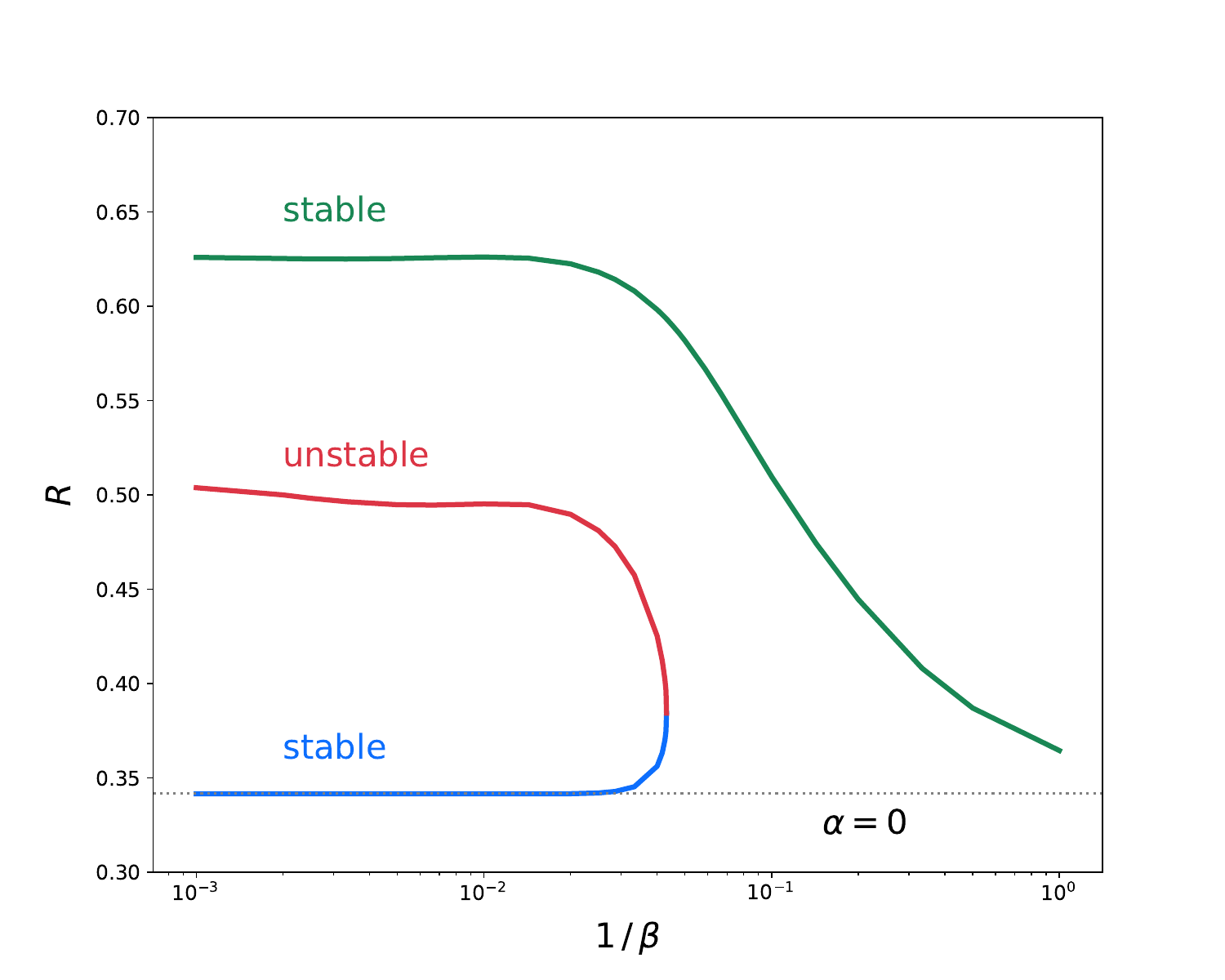}
\caption{
Bifurcation diagram showing the loci of the fixed points as functions of the inverse gate steepness (log scale).
Parameters are defined in the text.
The level corresponding to the stable point without score feedback ($\alpha = 0$) is also shown for reference (dotted line).
}
\label{bifurcation}
\end{figure}

The loci of the fixed points depend on the gate contrast and smoothness.
As shown in Figure \ref{bifurcation}, bistability is observed for sharp gates (large $\beta$).
As $\beta$ decreases, a stable + unstable pair annihilates leaving one stable branch, which further gradually evolves to the myopic limit.

\section{Concluding remarks}

This paper introduces a minimal two-tier stochastic control model for aggregator-routed RFQ markets in which a dealer's long-run win score feeds back into future RFQ opportunity intensity through a promotion gate.
The model separates the ``micro'' RFQ interaction -- size-ladder quoting with adverse selection and inventory risk -- from a ``macro'' routing layer in which the dealer is short-listed with a probability that depends on a slowly varying score.
By treating this score as an explicit state variable, the framework captures a practical feature of FX aggregator workflows: quote aggressiveness affects not only immediate edge capture but also future access to the most valuable stream of opportunities.

On the control side, we derived a reduced HJB in a Bergault-Gu\'eant operator form with an explicit win/lose branching structure for tier A, and standard non-local jump operators for the background tier.
An envelope-theorem argument yields a convenient representation of optimal controls: optimal win probabilities are given by derivatives of a one-dimensional reduced Hamiltonian, and optimal offsets follow by inversion of the win-probability curve.
This view helps interpret the policy in economic terms and provides a numerically robust route to recovering optimal quotes in multi-size RFQ ladders.

A key qualitative implication arises in the long-memory regime of the score (small EMA parameter).
In this limit, inventory control is fast and the score evolves on a slow manifold driven by the induced equilibrium win rate.
Combining a quadratic inventory ansatz with a local quadratic expansion of the RFQ Hamiltonians yields a tractable adiabatic closure: inventory curvature relaxes rapidly to a quasi-stationary level that depends on the total (tier A + tier B) liquidity available for inventory rebalancing, while the score dynamics reduce to a one-dimensional drift field.
Under a steep logistic promotion gate, this drift can undergo a fold bifurcation, generating bistability and hysteresis.
Economically, the optimal policy can alternate between ``campaigning'' (temporarily tighter quoting to move the score into a promoted basin) and ``harvesting'' (wider quoting while accepting gradual score decay), a pattern that is visible both in optimal offsets and in the instantaneous edge capture rate.

The two-tier structure plays an important stabilizing role.
Even when the promotion gate is near its minimum, the background stream maintains a baseline of inventory-mixing opportunities, preventing pathological blow-ups in the effective inventory curvature that can occur if all flow is gated by the score.
This feature is practically relevant: in real dealer franchises, not all client flow is routed through ranking-based shortlists, yet inventory risk is borne at the consolidated-book level.
The model highlights how ``residual'' flow can indirectly shape optimal behavior on the ranked stream by affecting the cost of inventory imbalances.

Several extensions are natural.
First, the gate $G(R)$ can be endogenized further by linking it to explicit top-list + exploration rules (finite-$N$ shortlisting) and to cross-sectional competition, moving beyond the reduced-form mean-field representation.
Second, the adverse selection component can be enriched to include post-trade informational drift or jump risk \cite{BarzykinBergaultGueantLemmel2025}, allowing one to study how informational asymmetry interacts with promotion incentives.
Third, the score itself could be multi-dimensional (wins, response times, hold times, reject rates) with potentially different platform weights.
Finally, from an empirical standpoint, the framework suggests a calibration program: estimate size-conditional win curves $p_z(\cdot)$, gate shapes $G(\cdot)$ and score memory parameters from RFQ logs, then test whether the predicted regions of campaigning/harvesting and the implied hysteresis align with observed regime switches in dealer quoting and flow.

\section*{Acknowledgment}
The author is grateful to Eric Mathew John (HSBC) for fruitful discussions and to Richard Anthony (HSBC) for support throughout the project and valuable comments.
The views expressed are those of the author and do not necessarily reflect the views or practices at HSBC.

\end{document}